\magnification=1200
\def\nonumfirst{\nopagenumbers
                \footline={\ifnum\count0=1\hfill
                           \else\hfill\folio\hfill
                           \fi}}
\nonumfirst
\magnification=1200
\def\singlespace{\baselineskip 12 pt}

\def\oneandahalfspace{\baselineskip 18pt}
\def\blankline{\vskip 12 pt\noindent}

\def\secto#1\endsecto{\vskip 20pt {\bf #1}\vskip 7pt\nobreak}
\global\newcount\refno \global\refno=1
\newwrite\rfile
\def\ref#1#2{\hbox{[\hskip 2pt\the\refno\hskip 2pt]}\nref#1{#2}}
\def\nref#1#2{\xdef#1{\hbox{[\hskip 2pt\the\refno\hskip 2pt]}}%
\ifnum\refno=1\immediate\openout\rfile=refs.tmp\fi%
\immediate\write\rfile{\noexpand\item{\noexpand#1\ }#2.}%
\global\advance\refno by1}
\def\semi{;\hfil\noexpand\break}
\def\demi{:\hfil\noexpand\break}

\newwrite\efile \let\firsteqn=T
\def\writeqno#1%
{\if T\firsteqn \immediate\openout\efile=eqns.tmp\global\let\firsteqn=F\fi%
\immediate\write\efile{#1 \string#1}\global\advance\meqno by1}

\def\eqnn#1{\xdef #1{(\the\secno.\the\meqno)}\writeqno#1}
\def\eqna#1{\xdef #1##1{(\the\secno.\the\meqno##1)}\writeqno{#1{}}}

\def\eqn#1#2{\xdef #1{(\the\secno.\the\meqno)}%
$$#2\eqno(\the\secno.\the\meqno)$$\writeqno#1}
\def\nobreak{\penalty1000}
\def\titl#1\endtitl{\par\vfil
     \vbox to 2in {}{\bf #1}\nobreak}
\def\titol#1\endtitol{\par\vfil
     \par\vbox to 1in {}{\bf #1}\par\vskip 1in\nobreak}
\def\tit#1\endtit{
     \vbox to 0.5in {}{\bf #1}\nobreak}

%
%
%
%
%
%
%
\def\lspace{\ifx\answ\bigans{}\else\qquad\fi}
\def\lbspace{\ifx\answ\bigans{}\else\hskip-.2in\fi} 
%
%
%

%
%
%
%

%

\def\bra#1{\left\langle #1\right|}
\def\ket#1{\left| #1\right\rangle}
\def\bk#1{\langle{\Psi _{#1}}|{\Phi _{#1}}\rangle}
\def\kb#1{\left|{\Phi _{#1}}\right\rangle \left\langle{\Psi _{#1}}\right|}
\def\kkb#1{\bigl|{{\rm K}_{#1}}\bigr\rangle \bigl\langle{{\rm K}^{\prime}_
{#1}}\bigr|}
\def\bbk#1{\langle{{\rm K}^{\prime}_ {#1}}|{{\rm K}_{#1}}\rangle}
\def\abs#1{\left| #1\right|}

%
%
\def\frac#1#2{{\textstyle{#1\over #2}}} 
%
%
%
%

%

%

\def\kl{ {\rm K_{L}} }
\def\ks{ {\rm K_{S}} }

\def\ksp{ {\rm K}^{\prime}_ {\rm S} }
\def\klp{ {\rm K}^{\prime}_ {\rm L} }

\def\gl{ {\it \Gamma}_{\rm L} }

%
%
%
\def\ltap{\ \raise.3ex\hbox{$<$\kern-.75em\lower1ex\hbox{$\sim$}}\ }
\def\gtap{\ \raise.3ex\hbox{$>$\kern-.75em\lower1ex\hbox{$\sim$}}\ }
\def\gl{\ \raise.5ex\hbox{$>$}\kern-.8em\lower.5ex\hbox{$<$}\ }
\def\roughly#1{\raise.3ex\hbox{$#1$\kern-.75em\lower1ex\hbox{$\sim$}}}
%
%

%
\def\[{\left[}
\def\]{\right]}
\def\({\left(}
\def\){\right)}
%
%

%
\textfont2=\tensy \scriptfont2=\sevensy \scriptscriptfont2=\fivesy
\def\cal{\fam2}
\def\Iscr{{\cal I}}

\def\Rscr{{\cal R}}
\def\Hscr{{\cal H}}
\def\Gscr{{\cal G}}

\def\Oscr{{\cal O}}
\def\Uscr{{\cal U}}
\def\Vscr{{\cal V}}
%

\def\st{\scriptstyle}

\def\pmb#1{\setbox0=\hbox{$#1$}%
  \kern-.025em\copy0\kern-\wd0
  \kern.05em\copy0\kern-\wd0
  \kern-.025em\raise.0433em\box0}
\def\pmbs#1{\setbox0=\hbox{$\st #1$}%
  \kern-.0175em\copy0\kern-\wd0
  \kern.035em\copy0\kern-\wd0
  \kern-.0175em\raise.0303em\box0}

\def\bfs#1{\hbox to .0035in{$\st#1$\hss}\hbox to .0035in{$\st#1$\hss}\st#1}



%

\def\bra#1{\langle #1 \vert }
\def\ket#1{\vert #1 \rangle }

%
%
\global\newcount\meqno \global\meqno=1
\newwrite\efile \let\firsteqn=T
\def\writeqno#1%
{\if T\firsteqn \immediate\openout\efile=eqns.tmp\global\let\firsteqn=F\fi%
\immediate\write\efile{#1 \string#1}\global\advance\meqno by1}

\def\eqqn#1#2{\xdef #1{(\the\meqno)}%
$$#2\eqno(\the\meqno)$$\writeqno#1}
\font\medf=cmb10 scaled \magstep3
%
\vsize=25 truecm
\hsize=16 truecm
\voffset=-0.8 truecm
%
\singlespace
\parskip 6truept
\parindent 20truept
\vbox{ {\rightline{\bf IFUM 571-FT/97}}
       {\rightline{\bf UNIBAS--TH 3/97}}
}
\hyphenation{ex-pe-ri-men-tal}
\hyphenation{va-cu-um}
\vskip 5truecm
\centerline{\medf 
                The Spectral Theory of Perturbative Decays}
\vskip 2truecm
\vskip 33truept
\centerline{D. Cocolicchio$^{(1,2)}$ and M. Viggiano$^{(1)}$}
\vskip 20truept
\vbox{
\centerline{\it $^{1)}$Dipartimento di Matematica,
Univ. Basilicata, Potenza, Italy}
\vskip 5truept
\centerline{\it Via N. Sauro 85, 85100 Potenza, Italy} }
\vskip 15truept
\vbox{
\centerline{\it $^{2)}$Istituto Nazionale di Fisica Nucleare,
                     Sezione di Milano, Italy}
\vskip 5truept
\centerline{\it Via G. Celoria 16, 20133 Milano, Italy} }
\vskip 4truecm
\centerline{\it ABSTRACT}
\vskip 15truept
\singlespace
\noindent
In this paper, we propose a complex approach to evaluate a function
sum of two noncommuting non Hermitian operators. Then, it is
proposed an explicit expansion of the evolution operator in the case
of the neutral $K$ meson system influenced by an external interaction.
Then, the importance of the procedure
is pointed out to consider the
algebraic expansion of the 
time evolution operator when ever
the dynamics decouples the internal transitions and center 
of mass motion.
\vskip 1truecm \noindent \singlespace
\vbox{
      {\leftline{IFUM 571-FT/97}}
      {\leftline{UNIBAS--TH 3/97}}
      }
\vfill\eject
\vsize=24 truecm
\hsize=16 truecm
\baselineskip 18 truept
\parindent=1cm
\parskip=8pt
\oneandahalfspace
%
%
\phantom{.}
\blankline
\leftline {\bf I. Introduction}
\blankline
\noindent
The temporal evolution of metastable systems is governed
by a non Hermitian Hamiltonian with non orthogonal eigenvectors
corresponding to complex eigenvalues. The problem to determine an
elegant and compact form for the evolution operator
\eqqn\uno{
\Uscr (t) = \exp[-i \Hscr t] \quad {\rm where} \quad \Hscr = \Hscr _0 
+ \Vscr}
is connected with the more general issue to express explicitly 
an arbitrary function of the sum of two noncommuting matrix operators.
One of the most tantalizing method to evaluate this matrix function
involves the subtleties of complex analysis and it was already
been developed in the particular case of Hermitian operators
\ref\PM{
P. Moretti and M. Mancini,
``{\it A Method of Calculating the Function of two
Noncommuting Operators}'', J. Math. Phys. {\bf 25} (1984) 2486}. 
The purpose of this paper is to extend this method to non Hermitian 
operators. The starting point is the generalized Cauchy's formula
\ref\EM{
E. Merzbacher,
``{\it Matrix Methods in Quantum Mechanics}'',
Am. J. Phys. {\bf 36} (1968) 814}
\eqqn\due{
f(A) = {1 \over 2 \pi i}\int\limits_\gamma f(z) \Gscr (z) \, dz
\quad ,}
here, the integral is extended over a contour $\gamma$ in the complex
$z$ plane which encloses all eigenvalues of $A$. It is then possible
to obtain an integral expression of $f(A)$ in terms of the well-known
resolvent operator
\eqqn\tre{
\Gscr_A (z) = (z\Iscr - A)^{-1} \quad .}
If $A$ and $B$ are two non Hermitian and noncommuting operators,
the resolvent operators of $A$ and of the sum $A+B$ are given,
respectively, by
\eqqn\quat{
\Gscr_A (z) = (z\Iscr - A)^{-1} \, , \; \;
\Gscr (z) = [z\Iscr - (A+B)]^{-1} \quad . }
In the convergence region of the geometric series
\eqqn\cinq{
{\left(1 -  {B \over z\Iscr - A}\right)}^{-1} = \sum_{n=0}^ \infty
{\left({B \over z\Iscr - A}\right)}^n \quad ,}
the following expansion for $\Gscr (z)$ holds
\eqqn\sei{
\eqalign{
\Gscr (z) = & {1 \over z\Iscr - A}\; {1 \over {z\Iscr - (A+B) \over
z\Iscr - A}} = {1 \over z\Iscr - A} \sum_{n=0}^ \infty
{\left({B \over z\Iscr - A}\right)}^n \cr
&  \cr
= & \sum_{n=0}^ \infty \Gscr _A (B \Gscr _A)^n =  
\sum_{n=0}^ \infty (\Gscr _A B)^n \Gscr _A \cr}}
which will be useful later. The right and left
eigenvectors of the operator $A$ are defined by the relations
\eqqn\sett{
A \ket{\Phi _i} = \lambda _i \ket{\Phi _i} \; , \quad
\bra {\Psi _i} A = \lambda _i \bra{\Psi _i}  \quad .}
Contrary to the case where $A$ is Hermitian, the sets
$\{\vert \Phi _i \rangle\}$ and $\{\langle \Psi _i \vert\}$, although
complete, they are not orthogonal and $\vert \Psi _i \rangle \ne 
\vert \Phi _i \rangle$. However, the following relation
\eqqn\ott{
\langle \Psi _i \vert \Phi _j \rangle =
\langle \Psi _i \vert \Phi _i \rangle \, \delta _{ij}}
holds, and therefore it is possible to generalize the
completeness relation using the following decomposition of unity
\eqqn\nov{
\Iscr = \sum_i^ {} {\kb{i} \over \bk{i}} \quad .}
We propose to give a spectral expansion of the function $f(A+B)$ in
terms of their relative eigenvalues. If $\Gamma$
is a closed contour enclosing the whole spectrum of the operator
$A+B$, then we have
\eqqn\ten{
f(A+B) = {1 \over 2 \pi i} \int\limits_\Gamma f(z) \Gscr (z) \, dz}
and from Eq. {\sei}
\eqqn\elev{
f(A+B) = {1 \over 2 \pi i}\sum_{n=0}^ \infty \int\limits_\Gamma
f(z) [(\Gscr _A B)^n \Gscr _A] \, dz \quad .}
The matrix elements will be obtained by
\eqqn\twel{
\eqalign{
& \bra{\Psi _1} f(A+B) \ket{\Phi _2} =
{1 \over 2 \pi i} \int\limits_\Gamma dz \, f(z) \sum_{n=0}^ \infty
\bra {\Psi _1} (\Gscr _A B)^n \Gscr _A \ket {\Phi _2} = \cr
&  \cr
& = {1 \over 2 \pi i}\sum_{n=0}^
\infty \int\limits_\Gamma dz \, f(z) \; \times \cr
& \times \; \left\{ \sum_{i_1,i_2,...,i_n}^ {}
\bra {\Psi _1} \Gscr _A B {\kb{i_1} \over \bk{i_1}} \Gscr _A B
{\kb{i_2} \over \bk{i_2}}...{\kb{i_n} \over \bk{i_n}}
\Gscr _A \ket{\Phi _2}\right\} = \cr
&   \cr
& = \sum_{n=0}^ \infty \sum_{\{i,n\}}^ {}  
\bra {\Psi _1} B {\kb{i_1} \over \bk{i_1}} B
{\kb{i_2} \over \bk{i_2}}...{\kb{i_{n-1}} \over \bk{i_{n-1}}}
B \ket{\Phi _2} \; \times \cr 
& \times \; {1 \over 2 \pi i} \int\limits_ \Gamma dz \, f(z)
\left[(z - \lambda _1)^{-1} (z - \lambda _{i_1})^{-1}...
(z - \lambda _{i_{n-1}})^{-1} (z - \lambda _2)^{-1} \right] \cr }}
where we have used the relation {\nov} and we have defined
$\{i,n\} \equiv \{i_1,i_2,...,i_{n-1}\}$. The indices $i_k$ run 
through the whole set of the eigenvectors as usual, whereas 
$\bra {\Psi _1}$ and $\ket {\Phi _2}$ are fixed.
If we introduce the following function
\eqqn\thirt{
F(z) = f(z) \left[(z - \lambda _1) (z - \lambda _{i_1})...
(z - \lambda _{i_{n-1}}) (z - \lambda _2) \right]^ {-1} \quad , }
we can write
\eqqn\fourte{
\eqalign{
& \bra{\Psi _1} f(A+B) \ket{\Phi _2} = \cr
& = \sum_{n=0}^ \infty \sum_{\{i,n\}}^ {}  
\bra {\Psi _1} B {\kb{i_1} \over \bk{i_1}} B
{\kb{i_2} \over \bk{i_2}}...{\kb{i_{n-1}} \over \bk{i_{n-1}}}
B \ket{\Phi _2} {1 \over 2 \pi i} \int\limits_\Gamma dz \, F(z) \quad ,
\cr }}
which generalize the result of the previous paper \PM.
Supposing that the eigenvalues of $A$ are enclosed within $\Gamma$, i.e. 
all the
singularities of the function $F(z)$ are inside the integral contour,
it is then possible to apply the theorem of the residues.
If we denote $\Rscr (\lambda _i)$ as the residue of $F(z)$
at the pole $z = \lambda _i$, the matrix
elements in the Eq. {\fourte} can be rewritten as
\blankline
\eqqn\fifth{
\eqalign{
\bra{\Psi _1} f(A+B) \ket{\Phi _2} = & \sum_{n=0}^ \infty
\sum_{\{i,n\}}^ {} \bra {\Psi _1} B {\kb{i_1} \over \bk{i_1}} B
{\kb{i_2} \over \bk{i_2}}...{\kb{i_{n-1}} \over \bk{i_{n-1}}}
B \ket{\Phi _2} \; \times \cr
\times & \; \left[ \Rscr (\lambda _1) + \sum_{\nu = 1}^ {n-1}
\Rscr (\lambda _{i_ \nu}) + \Rscr (\lambda _2) \right] \quad . \cr}}
If some eigenvalues are degenerate, the general expression of
$F(z)$ is
\eqqn\sixt{
F(z) =
f(z) \left[(z - \lambda _1)^ {m_1} (z - \lambda _{i_1})^ {m_{i_1}}
... (z - \lambda _{i_{n-1}})^ {m_{i_{n-1}}} (z - \lambda _2)^ {m_2}
\right]^ {-1}  }
where every exponent $m_i$ is the degeneracy order of the respective
eigenvalues $\lambda _i$. Finally we want to stress that the set
$\{m_i\}$ ($i \in \{1,...,n-1\}$) depends on the particular $\{i,n\}$
selected. It is worth noting that these results recover the formulae
already present in literature in the limiting case of the Hermitian
matrices, and they result generally more straightforward than the usual
algebraic methods
\ref\KF{K. O. Friedrichs, ``{\it Spectral Theory of Operators in
Hilbert Spaces}'', (Springer, New York, 1973);
T. Kato, ``{\it Perturbation Theory for Linear Operators}'', (Springer
Verlag, Berlin, 1976)}.
The use of these results can be displayed in the practical example
of the evolution operator.

\phantom{.}
\blankline
\leftline {\bf II. The evolution operator of the neutral kaon system}
\blankline
\noindent
The previous results let us make a decisive step toward a complete 
understanding of the controversial results about 
the dynamical behaviour of a decaying system described by the vector state
$\ket{\Psi (t)}$. Its time evolution  can be written by means of
an operator $\Uscr$:
\eqqn\sevent{
\ket{\Psi (t)} = \Uscr (t) \ket{\Psi (0)}  }
which can be expressed in the well-known exponential form, (using
units $\hbar = 1$)
\eqqn\eightt{
\Uscr (t) = \exp[-i \Hscr t] \quad .}
Although the Hamiltonian of a sensible quantum system is expected
to be a Hermitian operator, under suitable conditions we may recover
the time evolution according to an effective non Hermitian Hamiltonian
like in the case of metastable states. A celebrated example where
this description has proved extremely useful is the two-states kaon
complex. If this system is influenced by an external interaction, the
Hamiltonian operator can be written as
\eqqn\ninet{
\Hscr = \Hscr _0 + \Vscr }
where $\Hscr _0$ and the perturbation $\Vscr$ are, in general, two non
Hermitian and noncommuting operators. Thus, we can apply the formulas
of the previous section to expand $\Uscr (t)$ in terms of the eigenfunctions
$\ket{\ks}$, $\ket{\kl}$ of $\Hscr _0$
\eqqn\twen{
\eqalign{
\Hscr _0 \ket{\ks} = & \lambda _{\rm S} \ket{\ks} \cr
\Hscr _0 \ket{\kl} = & \lambda _{\rm L} \ket{\kl} \cr}}
where $\ket{\ks}$ and $\ket{\kl}$ are the right eigenvectors. The same
matrix operator $\Hscr _0$ has also two left eigenvectors with the same
eigenvalues
\eqqn\tone{
\eqalign{
\bra{{\rm K}^{\prime}_ {\rm S}} \Hscr _0 = & \lambda _{\rm S} \bra{{\rm K}^
{\prime}_ {\rm S}} \cr
\bra{{\rm K}^{\prime}_ {\rm L}} \Hscr _0 = & \lambda _{\rm L} \bra{{\rm K}^
{\prime}_ {\rm L}} 
\quad .\cr }}
The set 
$\{ \ket{{\rm K}^{\prime}_ {\rm S}}, \ket{{\rm K}^{\prime}_ {\rm L}} \}$ 
is the reciprocal
set of $\{ \ket \ks, \ket \kl \}$ in the sense that
\eqqn\ttwo{
\langle {\rm K}^{\prime}_ {\rm S} \vert \kl \rangle = 0 =
\langle {\rm K}^{\prime}_ {\rm L} \vert \ks \rangle \quad .}
If we normalize all the eigenvectors to 1 and denote the overlap as
\eqqn\ttrh{
\chi = \langle \kl \vert \ks \rangle}
then
\eqqn\tfour{
\eqalign{
\ket{\ksp} = & {1 \over \sqrt{1 - {\abs \chi}^2}} (\ket\ks -
\chi \ket \kl) \cr
&    \cr
\ket{\klp} = & {1 \over \sqrt{1 - {\abs \chi}^2}} (\ket\kl -
\chi^ * \ket \ks) \cr}}
where
\eqqn\tfive{
\langle \ksp \vert \ks \rangle = \sqrt{1 - {\abs \chi}^2}
= \langle \klp \vert \kl \rangle }
and
\eqqn\tsix{
\langle \klp \vert \ksp \rangle = -\chi
\quad .}
Therefore, we have the following equivalent decomposition of unity
\blankline
\eqqn\tseve{
\eqalign{
\Iscr = & \ket \ks \bra \ks + \ket \klp \bra \klp = \cr
&   \cr
= & \ket \kl \bra \kl + \ket \ksp \bra \ksp = \cr
&   \cr
= & {1 \over \sqrt{1 - {\abs \chi}^2}} \; \left(
\ket \ks \bra \ksp + \ket \kl \bra \klp
\right) = \cr
&   \cr
= & {1 \over \sqrt{1 - {\abs \chi}^2}} \; (\ket \ksp \bra \ks
+ \ket \klp \bra \kl) \quad .\cr} }
In the particular case of the evolution operator, the generalized
Cauchy's formula is written in the form
\eqqn\teigh{
\Uscr (t) = {1 \over 2 \pi i} \int\limits_\Gamma e^{-izt} \Gscr (z) \, dz}
where the resolvent operator is
\eqqn\tnine{
\Gscr (z) =
{1 \over z\Iscr - \Hscr} \; = \; {1 \over z\Iscr - [\Hscr_0 + \Vscr]}}
and $\Gamma$ is a closed curve encircling all the complex eigenvalues
of the total Hamiltonian $\Hscr$. If
\eqqn\thirty{
\Gscr _0 (z) = {1 \over z\Iscr - \Hscr_0}}
is the resolvent operator of $\Hscr_0$, an analogous expansion to Eq.{\sei} 
for $\Gscr (z)$ holds by substituting $\Hscr_0$ and $\Vscr$ for the operators $A$
and $B$ respectively. Now, the matrix elements will be obtained by
\eqqn\thone{
U_{\alpha \, \beta} = \bra {{\rm K}^{\prime}_ \alpha} \exp[-i \Hscr t]
\ket {{\rm K}_\beta}}
where the greek letters $\alpha$ and $\beta$ are fixed and
$\alpha , \beta \in \{{\rm {S,L}}\}$. From Eq.{\twel} we have
\blankline
\eqqn\thtwo{
\eqalign{
& \bra {{\rm K}^{\prime}_ \alpha} \exp[-i \Hscr t]\ket {{\rm K}_\beta}= \cr
&   \cr
& = \sum_{n=0}^ \infty \sum_{\{\mu,n\}}^ {}  
\bra {{\rm K}^{\prime}_ \alpha} \, \Vscr \, {\kkb{\mu_1} \over \bbk{\mu_1}} 
\, \Vscr \,
{\kkb{\mu_2} \over \bbk{\mu_2}}\, ...\, {\kkb{\mu_{n-1}} \over \bbk{\mu_{n-1}}}
\, \Vscr \, \ket{{\rm K} _\beta} \; \times \cr 
& \times \; {1 \over 2 \pi i} \int\limits_ \Gamma dz \, e^{-izt}
\left[(z - \lambda _\alpha)^{-1} (z - \lambda _{\mu_1})^{-1}...
(z - \lambda _{\mu_{n-1}})^{-1} (z - \lambda _\beta)^{-1} \right]= \cr
&  \cr
& = \sum_{n=0}^ \infty \sum_{\{\mu,n\}}^ {}  
\bra {{\rm K}^{\prime}_ \alpha} \, \Vscr \,{\kkb{\mu_1} \over \bbk{\mu_1}} 
\, \Vscr \, {\kkb{\mu_2} \over \bbk{\mu_2}}...{\kkb{\mu_{n-1}} \over
\bbk{\mu_{n-1}}} \, \Vscr \, \ket{{\rm K} _\beta} \times \cr
& \times \; \left[ \Rscr (\lambda _\alpha) + \sum_{\nu = 1}^ {n-1}
\Rscr (\lambda _{\mu_ \nu}) + \Rscr (\lambda _\beta) \right] \quad . \cr}}
\blankline
Here $\{\mu,n\} \equiv \{\mu_1, \mu_2,..., \mu_{n-1}\}$ and the indices
$\mu_k$ are varying in the set $\{\rm {S,L}\}$. We can give a more explicit
expression to the quantity
$\bra {{\rm K}^{\prime}_ \alpha} \exp[-i \Hscr t]
\ket {{\rm K}_\beta}$ with the use of formulas {\ttrh, \tfour, \tfive}
\eqqn\thtrh{
\eqalign{
& U_{\alpha \, \beta} = \bra {{\rm K}^{\prime}_ \alpha} \exp[-i \Hscr t]
\ket {{\rm K}_\beta}= \cr
&  \cr
& = \sum_{n=0}^ \infty {1 \over \left[\sqrt{1 -{\abs \chi}^2}\right]^ 
{(n-1)}}
\bra {{\rm K}^{\prime}_ \alpha}  \, \Vscr \, \bigl [ \;
\ket \ks \bra \ksp + \ket \kl \bra \klp \;
\bigr ] \, \Vscr \, \cr
&  \cr
& \bigl [ \;
\ket \ks \bra \ksp + \ket \kl \bra \klp  \;
\bigr ]... \bigl [ \;
\ket \ks \bra \ksp + \ket \kl \bra \klp  \;
\bigr ] \,
\Vscr \, \ket{{\rm K} _\beta} \; \times \cr
&  \cr
& \times \; \bigl [ \; \Rscr (\lambda _\alpha) + \sum_{\nu = 1}^ {n-1}
\Rscr (\lambda _{\mu_ \nu}) + \Rscr (\lambda _\beta) \bigr ] \; \quad .
\cr}}
Being $\lambda_ \alpha$, $\lambda_ \beta$ and $\lambda_ {\mu_ \nu}$
($\nu \in \{1,2,...,n-1\}$) equal to $\lambda_ {\rm S}$ or $\lambda_ {\rm L}$,
Eq.~Ê\thtrh will be therefore rewritten as
\eqqn\thfou{
\eqalign{
U_{\alpha \, \beta} = & 
\sum_{n=0}^ \infty {1 \over \left [\sqrt{1 -{\abs \chi}^2}\right]^ {(n-1)}}
\bra {{\rm K}^{\prime}_ \alpha} \, \Vscr \, \bigl [ \;
\ket \ks \bra \ksp + \ket \kl \bra \klp  \;
\bigr ] \, \Vscr \, \cr
&  \cr
& \bigl [ \;
\ket \ks \bra \ksp + \ket \kl \bra \klp  \;
\bigr ]... \bigl [  \;
\ket \ks \bra \ksp + \ket \kl \bra \klp \;
\bigr ] \,
\Vscr \, \ket{{\rm K} _\beta}
\bigl [ \; \Rscr (\lambda_ {\rm S}) + \Rscr (\lambda_ {\rm L}) \; \bigr ] 
\quad , \cr}}
where $\Rscr (\lambda_ i)$ is the residue at $z=\lambda_ i$, $\lambda_ i
\in \{ \lambda_ {\rm S}, \lambda_ {\rm L} \}$, of the function
\eqqn\thfive{
F(z) = e^{-izt} \bigl [\; (z - \lambda_ {\rm S})^r (z - \lambda_ {\rm L})^s \; 
\bigr ]^{-1} }
and $r$ and $s$ are positive integer numbers subject to the condition
$r + s = n+1$. In a general theory
\eqqn\thsix{
\lambda_ {\rm {S,L}} = {{\rm {tr}} \, \Hscr _0 \; \mp \;
\sqrt{[ \, {\rm {tr}} \, \Hscr _0 \,]^2 - 4 {\rm {det}} \, \Hscr _0}
\over 2}}
so $\lambda_ {\rm S} \ne \lambda_ {\rm L}$ and all eigenvalues are not 
degenerate.
In this way if $n>2$,
$\lambda_ {\rm S}$ and $\lambda_ {\rm L}$ are not simple poles 
for $F(z)$ and a direct calculation of residues gives
\eqqn\thsev{
\Rscr (\lambda_ {\rm S}) = {1 \over (r-1)!} \; {d^{r-1} \over dz^{r-1}} \bigl
[ \, e^{-izt} (z - \lambda_ {\rm S})^{-s} \, \bigr ] \bigg |_ {z= \lambda_ 
{\rm S}}}
and
\eqqn\theig{
\Rscr (\lambda_ {\rm L}) = {1 \over (s-1)!} \; {d^{s-1} \over dz^{s-1}} \bigl
[ \, e^{-izt} (z - \lambda_ {\rm L})^{-r} \, \bigr ] \bigg |_ {z= \lambda_ 
{\rm L}}
\, . }
We display the use of these formulas to calculate the second order
term in expansion in the particular case $\alpha = {\rm S}$ and 
$\beta = {\rm L}$
\eqqn\thnin{
\eqalign{
U^{(2)}_ {{\rm {S,L}}} = & {1 \over \sqrt {1 - {\abs {\chi}}^2}} \biggl [
\; \bra \ksp \, \Vscr \, \ket \ks \bra \ksp \, \Vscr \, 
\ket \kl \;
[ \, \Rscr _{(2)} (\lambda_ {\rm S}) + \Rscr _{(1)} (\lambda_ {\rm L}) \, ] \; 
+ \cr
+ & \bra \ksp \, \Vscr \, \ket \kl \bra \klp \, \Vscr \, \ket \kl \;
[ \, \Rscr _{(1)} (\lambda_ {\rm S}) + \Rscr _{(2)} (\lambda_ {\rm L}) \, ] \;
\biggr ]
= \cr
&   \cr
= & {1 \over \sqrt {1 - {\abs {\chi}}^2}}  \;
\bra \ksp \, \Vscr \, \ket \ks \bra \ksp \, \Vscr \, \ket \kl \; \biggl [ \,
e^{-i \lambda_ {\rm S} t} \left[ \, (\lambda_ {\rm L} - \lambda_ {\rm S})^{-1} 
\, + \,
(\lambda_ {\rm L} - \lambda_ {\rm S})^{-2} \, \right ] \; + \cr
+ & \; e^{-i \lambda_ {\rm L} t} (\lambda_ {\rm L} - \lambda_ {\rm S})^{-2} 
\, \biggr ] \; +\cr 
+ & {1 \over \sqrt {1 - {\abs {\chi}}^2}} \;
\bra \ksp \, \Vscr \, \ket \kl \bra \klp \, \Vscr \,\ket \kl \; \biggl [ \,
e^{-i \lambda_ {\rm L} t} \left[ \, (\lambda_ {\rm S} - \lambda_ {\rm L})^{-1} 
\, + \,
(\lambda_ {\rm S} - \lambda_ {\rm L})^{-2} \, \right ] \; + \cr
+ & \; e^{-i \lambda_ {\rm S} t} (\lambda_ {\rm S} - \lambda_ {\rm L})^{-2} 
\, \biggr ]   \; .\cr
}}
Here we have labelled $\Rscr (\lambda_ i)$ with the subscripts $(1)$ and $(2)$
to stress that $\lambda_ i$ is a first or a second order pole. Obviously,
the presence of higher order terms involves higher order poles, 
but, in the case of the neutral $K$ meson system, the expression of $U_ {\alpha
\beta}$ is not so cumbersome as in the general case. In view of this
consideration, this perturbative approach is extremely useful in the
description of evolution of the two-states kaon complex
\ref\CDMV{D.Cocolicchio and M. Viggiano,
``{\it The Quantum Theory of the Kaon Oscillations}'',
preprint IFUM FT 97, Istituto Nazionale di Fisica Nucleare, Sezione di Milano}. 
But, as mentioned
in \PM, the use of the complex analysis is more convenient and it can be 
successfully applied also when the usual time-dependent
perturbation theory fails.

\blankline
\leftline {\bf III. Concluding Remarks}
\blankline
\noindent
Outside the realm of particle physics, there are many other cases of
unstable systems influenced by external interactions, 
where the previous approach
becomes indispensible. For example, in modern quantum optics,
it seems particularly important to analyze the (para)magnetic resonance 
\ref\RRS{I.
I. Raby, N. F. Ramsey and J. Schwinger,
``{\it  Use of Rotating Coordinates in Magnetic Resonance Problems}'',
Rev. Mod. Phys. {\bf 26} (1954)  167;
R. P.
Feynmann, F. L. Vernon and R. W. Hellworth, 
``{\it Geometrical Representation of the Schrodinger Equation for 
Solving Maser Problems}'',
J. Appl. Phys. {\bf 28} (1957) 
49},
and in general to describe the two states (spin-up, spin-down) involving
electrons and protons with dissipation.
Presently, it provides the theoretical framework to study
a multitude of effects involving laser dynamics.
Nevertheless, unstable two level systems in interaction with other degrees of
freedom require the strategy outlined before.
The system is, in fact, an open system and its dynamical behaviour under the
influence of an external interaction can be described only redefining the
evolution operator
\eqqn\utrafo{
\Oscr (t) \Uscr(t) \Oscr^{-1}(0)= \tilde{\Uscr}(t) =
   \exp [-i t \tilde{\Hscr}/\hbar ] = e^{A+B} \quad . }
The transformation
$\Oscr (t)$ decouples the internal degrees of freedom from the
motion of the center of mass and provides a time independent 
Hamiltonian $\tilde\Hscr
 = \Oscr \Hscr \Oscr^{-1} -i \hbar \Oscr {\dot \Oscr}^{-1}$. 
The operators $A$ and $B$ are introduced to clarify the
mathematical structure of the calculation below.
The new time evolution operator $\tilde \Uscr (t)$ is then determined by 
the exponential factorization of two non commuting (sometimes non 
Hermitian) operators. It is evident now,
the importance of the method outlined above
which turns out to be particularly compelling as far as the physical
interpretation is concerned.
It is worth discussing in connection with the algebraic approach.
This method makes use of the parametric differentiation
of the exponential of an operator and of the commutation
relations in the context of the Baker-Campbell-Hausdorff
(BCH) formula
\ref\LUT{ R. M. Wilcox, 
``{\it Exponential Operators and Parameter Differentiation in Quantum 
Physics}'', J. Math. Phys. {\bf 8} (1967) 962;
M. Lutzky, 
``{\it Parameter Differentiation of Exponential Operators and the
Baker - Campbell - Hausdorff Formula}'',
J. Math. Phys. {\bf 9} (1968) 1125;
J. A. Oteo, ``{\it The Baker - Campbell - Hausdorff
formula and nested commutator identities}'', J. Math. Phys.
{\bf 32} (1991) 419}.
Then we can think to
separate the center of mass $A$ part evolution to
factorize $\tilde{\Uscr}(t)$ according to
\eqqn\ut{
    \tilde{\Uscr}(t) = e^A W(t) \; .}
Thus the complete time evolution of the two-level system is then based 
on the remaining determination of the operator $W(t)$ which contains
the influence of the external interactions on the internal dynamics. 
On this point, to work out $W (t)$ of Eq.~\ut\ 
we consider  the operator
\eqqn\ansatz{
   \Gscr(\lambda) = \exp [ \lambda (A+B)] = e^{\lambda A}
   W(\lambda) \; ,}
and restrict to $\lambda=1$ at the end. Differentiation
of Eq.~\ansatz\  with respect to $\lambda$ leads to the
differential equation
\eqqn\dglu{
   \frac{dW}{d \lambda} = \left( e^{-\lambda A}B\; e^{\lambda A}
   \right ) W(\lambda) 
\simeq  \left ( B - \lambda [A,B] 
   \right ) W(\lambda) }
with the initial condition $W(\lambda=0) = {\bf 1}$ and
using the identity
\eqqn\anoni{
   e^{-\lambda A} B\; e^{\lambda A} = \sum_{n=0}^\infty
   \frac{\lambda^n}{n!} K_n \quad ,\quad K_0 = B \; , \;
   K_{n+1} = [K_n, A] }
which holds for any two operators $A,B$.
Under general assumptions, it
may be written as a matrix equation of the form,
\eqqn\anond{
\left(\matrix{
dW_{11}/d \lambda & dW_{12}/d\lambda \cr
dW_{21}/d \lambda & dW_{22}/d\lambda \cr} \right) 
= 
\left(\matrix{
P - \lambda [Q,P] & R \cr
R & -P + \lambda [Q,P] \cr} \right)
\left(\matrix{
W_{11} & W_{12} \cr
W_{21} & W_{22} \cr} \right )\; . }
This result assumes that the factorization is able to select the non 
vanishing commutator $\left[ A, B\right] =\left[ Q, P\right]$, that 
is, however, a c-number, whereas $R$ is the remaining 
part in $B$. An exemplification is represented by
\eqqn\aru{\eqalign{
A =& Q {\bf 1} \cr
B =& P {\bf \sigma}_3 + R  {\bf \sigma}_1 \; . \cr} }
Eq.~\anond\ is an operator-valued system of differential equations, but
it contains only commuting operators
so that we can treat it as an ordinary differential
equation with the initial condition  $W_{ij}(\lambda=0)=1$.

Inserting the equation for $dW_{11}/d \lambda$
into the equation for $dW_{21}/d \lambda$
and similarly for
$dW_{22}/d \lambda$ into that for $dW_{12}/d \lambda$, one gets
\eqqn\dglt{\eqalign{
     \frac{d^2 W_{11}}{d \lambda^2} =& \{ R^2 - \left[Q,P\right] +
     (\lambda  \left[Q,P\right] - P)^2 \} W_{11}  \cr
     \frac{d^2 W_{22}}{d \lambda^2} =& \{ R^2 +\left[Q,P\right] +
     (\lambda \left[Q,P\right] - P)^2 \} W_{22} \; .\cr} }
After the introduction of the parameter
\eqqn\anot{
\theta = \frac{R^2}{2[Q,P]} }
and the change of the variable $y = (\lambda [Q,P] -P)
\sqrt{2/[Q,P]}$, Eq.~\dglt\  becomes
\eqqn\dglii{
\eqalign{
     \frac{d^2 W_{11}}{d y^2} =& \left \{ \frac{y^2}{4} +
     \theta - {1\over 2} \right \} W_{11}(y) \cr
     \frac{d^2 W_{22}}{d y^2} =& \left \{ \frac{y^2}{4} +
     \theta + {1\over 2} \right \} W_{22}(y) \cr} }
with the initial conditions
$W_{11}(\lambda=0)=W_{22}(\lambda=0)=1$ and
\eqqn\anonc{
      \frac{dW_{11}}{d y}\Bigg |_{\lambda=0} = - \frac{dW_{22}}{dy}
      \Bigg |_{\lambda=0} =
      {P\over 2} \sqrt{\frac{2}{[Q,P]}}   \; .}
The solution of Eqs.~\dglii\  is a
linear combination of
parabolic cylinder functions.
The total operator $\tilde{\Uscr}(t)$ is given by
\eqqn\lsg{
\tilde{\Uscr}(t) = \exp \left \{ A \right \}
\left(
\matrix{
W_{11} & W_{12} \cr
W_{21} & W_{22} \cr} \right) \; .} 
It remains to cancel the initial unitary transformation $\Oscr (t)$
in Eq.~\utrafo\  to obtain the exact expression for the
time evolution operator:
\eqqn\trlsg{
     \Uscr (t) = \Oscr ^{-1}(t)\,  \tilde{\Uscr}(t)\,  \Oscr(0) 
          = \exp \left \{ A \right \}
\Oscr ^{-1}(t)\,  W (t)\,  \Oscr(0) \; .}
An instructive consistency check is to turn off the external
interaction by setting $\theta$ to zero. In this case the
stable two-level system should be recovered. $\theta =0$
implies immediately
\eqqn\anosi{
     \Uscr(t) = \exp \left \{ {-it} \Hscr _{cm}\right \} 
\exp \left \{ {-it} \Vscr \right \} }
as it was to be expected. 
The first term describes the free motion
of the system, and the second term contains the internal
transitions.
It is also possible to derive an
expansion for small $\theta$. But 
the result is then difficult to understand since
it contains various combinations of error functions.

\noindent
In this paper, we have analyzed the dynamical evolution of unstable systems under
the influence of external interactions. The generalization of
the complex spectral theory is proposed to account for these unstable open systems.
Furthermore, the results of the coupled dynamics of the internal transitions and
the center of mass motion are worked out with the algebraic expansion of the time
evolution operator.

\blankline
\bigskip
\centerline{\bf ACKNOWLEDGMENTS}
\bigskip
\noindent
One of us (M.~ÊV.) wishes to thank the warm hospitality of
the Department of Mathematics of the University of Basilicata, Potenza.
\bigskip

\vfil\eject
\vfill\eject\immediate\closeout\rfile
\centerline{{\bf References}}\bigskip
\input refs.tmp\vfill\eject
 \bye